\documentclass{appolb}
\begin{document}


\title{SELECTED ASPECTS OF PHYSICS OF FERMIONIC BUBBLES%
\thanks{Presented at the {\it High Spin Physics 2001} NATO
Advanced Research Workshop, dedicated to the memory of Zdzis\l{}aw Szyma\'nski,
Warsaw, Poland, February 6--10, 2001
}%
}

\author{Piotr Magierski$^{1}$ and Aurel Bulgac$^{2}$
\address{
$^{1}$Institute of Physics, Warsaw University of Technology, 
Warsaw, POLAND \\
$^{2}$Department of Physics,  University of Washington,
Seattle, USA }
}
\date{May 17, 2001}
\maketitle

\begin{abstract}
We discuss properties of the Fermi system which contain
one or more spherical (or almost spherical) objects.
The interplay between
various effects, such as shell correction and chaotic behavior
is considered. We briefly mention
the role of the temperature, pairing, and effects associated
with bubble dynamics.
\end{abstract}

\PACS{ 21.10.Dr, 21.65.+f, 97.60.Jd, 67.55.Lf}

\bigskip

The term ``fermionic bubble'' is used here to denote an (almost) spherical
impurity/inhomogeneity
immersed in an otherwise homogeneous fermionic system.
There are a number of situations where such systems can
be formed. In particular, halo nuclei \cite{austin},
bubble nuclei \cite{yu},
highly charged alkali metal clusters \cite{dietrich},
various heterogeneous
atomic clusters \cite{saito} and
neutron star crust \cite{bul} are a few scattered examples of
systems which may be regarded as containing bubbles.
Fermions reside there in a rather
smoothly behaving mean--field potential, except for the regions where bubbles
are formed and where it changes its depth. A simple semiclassical
analysis shows that if the difference in depths
is large then the amplitude
of fermionic wave function
inside the bubble is small (in most situations) and consequently
a bubble will act approximately like a hard wall.

Despite the fact that conditions that lead to
formation of fermionic bubbles as well as their stability
depend on a particular
system under consideration, there are many aspects
of bubble physics which are generic.
For example,
in the case of a finite one bubble system the question
arises concerning the most favorable position of the bubble
inside the system. It is tacitly assumed that
bubble position has to be determined according to symmetry
considerations. For a Bose condensate one can easily show that a bubble
has to be off--center \cite{chin} but in the case of a Fermi system the
most favorable arrangement is not obvious \cite{yu,bubbles}.
The total energy of a many fermion system has the general form:
%--------------------------------------------------------------------
\begin{equation}
E(N)=e_v N +e_s N^{2/3}+e_c N^{1/3} + E_{sc}(N),
\label{eq:liq}
\end{equation}
%--------------------------------------------------------------------
where the first three terms represent the smooth liquid drop part of
the total energy and $E_{sc}$ is the pure quantum shell correction
contribution, the amplitude of which grows in magnitude approximately
as $\propto N^{1/6}$, see Ref. \cite{strutin}.
The character of the shell correction is in general
strongly correlated with the existence of regular and/or chaotic
motion \cite{balian,strut}. If a spherical bubble appears in a
spherical system and if the bubble is positioned at the center, then
for certain ``magic'' fermion numbers the shell correction energy
$E_{sc}(N)$, and hence the total energy $E(N)$, has a very deep
minimum. However, if the number of particles is not ``magic'', in
order to become more stable the system will in general tend to deform.
This situation is very similar to the celebrated
Jahn--Teller effect in molecules. One has to remember however
that real deformations lead to an increased liquid drop
energy, whereas merely shifting a bubble off--center
deforms neither the bubble nor the external surface and therefore
the liquid drop part of the total energy for neutral systems
remains unchanged.
Thus it is expected that the one-bubble system, for ``non-magic''
number of fermions, will rather exhibit
softness toward the off-center displacements of the bubble.
On the other hand as the bubble is moved
off center, the classical problem becomes more chaotic \cite{bohigas}
and one can naively expect that the single particle (s.p.)
spectrum would approach that of a
random Hamiltonian \cite{gian}, and that the nearest--neighbor
splitting distribution would be given by the Wigner
surmise\cite{mehta}. This is however not the case and
one can show \cite{yu} that even for extreme
displacements large gaps in the s.p. spectrum occur significantly more
frequently than in the case of a random (which is closer to an uniform)
spectrum. One should however keep in mind that large gaps in the
energy spectra can occur if several noninteracting chaotic spectra are
superimposed as well \cite{hans}.

The formation of two or more bubbles at
the same time, in finite, infinite or semi--infinite systems
opens a plethora of new problems associated with their
mutual interactions and the most favorable positional arrangements.
For the sake of simplicity
let us consider first two spherical, identical bubbles
that have been formed in an
otherwise homogeneous and
electrically neutral fermionic matter. We
shall also assume that the bubbles are completely hollow,
stable and rigid.
According to a liquid drop model approach
the energy of the system should be insensitive to the relative
positioning of the two bubbles.
In the semiclassical approach, which is justified for the
''sizeable'' bubbles (i.e. when the Fermi wavelength is small
comparing to the size of the bubble), the shell correction energy is
determined by periodic orbits in the system.  In the case of two
spherical bubbles there exist only one such trajectory (with
repetitions) which gives rise to the interaction energy between
bubbles \cite{bul}. The interaction energy
between the two bubbles of radii $R$,
due to the existence of the periodic orbit
at large separations $d$, reads:
\begin{equation} \label{shell}
E_{shell}=\frac{1}{8\pi}\frac{\hbar^2 k_F }{m}\frac{R^2}{d^3}\cos
(2k_{F}d),
\end{equation}
where $k_{F}$ denotes the Fermi momentum and $m$ the fermion mass.
This semiclassical formula approximates the exact result
surprisingly well even for relatively small distances \cite{bulwirz}.
Although the above result resembles the
well known Ruderman--Kittel interaction between two point--like
impurities, the interaction
(\ref{shell}) is much stronger since the large obstacle reflects
back more of the incident wave than a point object, which acts
like a pure $s$--wave scatterer.

The interaction between many bubbles in the Fermi system
may look, at the first glance, quite complicated
since three--, four--, and other
many--body terms will appear as a result
of multiple fermion scattering
between bubbles. One can show however that
many-body terms are quite small and
give merely a small corrections to the dominant pairwise
interaction \cite{bulwirz}.
If the bubble density in the system becomes sufficiently large
then a new form of matter can be created,
foam. One might argue that sometimes a ``misty'' state could be more
likely. As in the case of percolation, whether a ``foamy'' or a
``misty'' state would be formed, should strongly depend on the average
matter density. At very low average densities, formation of droplets
is more likely, while at higher average densities (lower than the
equilibrium density however) the formation of a foam is more
probable.

The best example of the many--bubble system is the neutron star
where various inhomegeneities in the neutron matter may be formed.
Apparently an agreement has been reached
concerning the existence of the following chain of phase changes as
the density is increasing: nuclei $\rightarrow$ rods $\rightarrow$
plates $\rightarrow$ tubes $\rightarrow$ bubbles $\rightarrow$ uniform
matter.  The density range for these phase transitions is $0.04 - 0.1
fm^{-3}$ \cite{lor,chris}. At
densities of the order of several nuclear densities the quark degrees
of freedom get unlocked and the formation of various quark matter
droplets embedded in nuclear matter becomes then energetically
favorable \cite{heiselberg}.
The appearance of different phases is attributed to the interplay
between the Coulomb and surface energies. However in the phase
transition region the relative energies between phases coming
from the liquid drop model are small and the shell effects
coming from the interaction of the type (\ref{shell})
starts
to play a dominant role
\cite{bul}. Our
results suggest that the inhomogeneous phase has perhaps an
extremely complicated structure, maybe even completely disordered,
with several types of shapes present at the same time.

All the fermionic system possessing bubbles are usually characterized
by large pairing interaction which we neglected so far in our
considerations. One should remember however that pairing
correlations will be significant when the Fermi level occurs in a
region of high s.p. level density and ultimately
lead to a leveling of the potential energy surface.
Also with increasing temperature the shell correction energy decreases.
Certainly for large temperatures the positional entropy will
determine the most favourable positions of the bubbles which
for the one-bubble system will be still off-center\cite{yu}.
At intermediate temperature $T$ the free energy associated with
the bubble-bubble interaction is given in the form:
\begin{equation}
F_{shell}=-T\int_{0}^{\infty}
[ g(\varepsilon ,{\bf l} )-g_0(\varepsilon ) ]
\log\left [ 1+\exp \left ( -\frac{\epsilon-\mu}{T}\right )\right ] d\epsilon ,
\end{equation}
where $g_0 (\varepsilon )$ is the density of states 
with objects infinitely separated, $g (\varepsilon ,{\bf
l})$ is the density of states in the presence of
bubbles and ${\bf l}$ is an ensemble of geometrical parameters
describing these objects and their relative geometrical
arrangement.
One can show that it leads to an exponential suppression
of the interaction strength
as compared to eq. (\ref{shell}) \cite{bul2}.

The energetics of two or more bubbles, their relative placements and
positions with respect to boundaries, their collisions and bound state
formation, their impact on the role played by periodic or chaotic
trajectories, and their temperature dependence, are but a few in a
long list of challenging questions.  A plethora of new, extremely soft
collective modes is thus generated.  The character of the response of
such systems to various external fields is an extremely intricate
issue.  Since the energy of the system changes only very little while
the bubble(s) is being moved, a slight change in energy can result in
large scale bubble motion. However still the problem of moving bubbles is
far from being fully understood.
One expects that a bare bubble mass $M$
should be renormalized in the same way as a bare mass of a
particle is renormalized in the quantum field theory, since the impurity
travels in the fermionic medium in a cloud of medium excitations.
Since the mass is large comparing to fermion masses
the motion of the bubble will be governed by two main physical processes:
an infrared divergence leading to Anderson's orthogonality catastrophe
and recoil of the bubble. The former effect dominate
for $M\rightarrow\infty$ and the latter for $M\rightarrow 0$ (see e.g.
\cite{rosch} and references therein).
However there is no agreement so far on the relative importance
of these two effects in the general case.

Hopefully, the few results highlighted here will convince the reader
of the richness of these systems where both static and dynamic properties
are challenging to describe.

\bigskip
We appreciate the financial support of DOE and
of the Polish Committee for Scientific Research (KBN) and many discussions
with our collaborators A. Wirzba, Y. Yu, S.A. Chin and H. Forbert.


\begin{thebibliography}{99}

\bibitem{austin} S.M. Austin and G.F. Bertsch, Scientific American,
 {\bf 272}, 62 (1995).

\bibitem{yu} Y. Yu, A. Bulgac and P. Magierski, Phys. Rev. Lett.
 {\bf 84}, 412 (2000); J.A. Wheeler, unpublished notes;
 C.Y. Wong, Ann. Phys. {\bf 77}, 279 (1973).

\bibitem{dietrich} K. Pomorski and K. Dietrich, Eur. Journ. Phys.
 {\bf D 4}, 353 (1998).

\bibitem{saito} S. Saito and F. Yabe, in {\it
Chemistry and Physics of Fullerenes and Related Materials}, vol. {\bf
6}, eds. K.M. Kadish and R.S. Ruoff, Pennington, 1998, pp 8--20; T.P. Martin
{\it et al.}, J. Chem. Phys. {\bf 99}, 4210 (1993); U. Zimmermann {\it
et al.}, Phys. Rev. Lett.  {\bf 72}, 3542 (1994).

\bibitem{bul} A. Bulgac, P. Magierski, Nucl. Phys. {\bf A683} (2001) 695;
Phys. Scripta {\bf T90} (2001) 150; Acta Phys. Polon. {\bf B32} (2001) 1099.

\bibitem{chin} S.A. Chin and H.A. Forbert, Phys. Lett. {\bf A272}, 402 (2000).

\bibitem{bubbles} A. Bulgac, S.A. Chin, H. Forbert, P. Magierski and Y. Yu,
   in Proc. of the Int. Workshop on {\it Collective excitations in Fermi and
   Bose systems''}, eds. C.A. Bertulani, L.F. Canto and M.S. Hussein, pp.
   44--61, World Scientific, Singapore (1999).

\bibitem{strutin} V.M. Strutinsky and A.~G. Magner,
Sov. J. Part. Nucl. Phys. {\bf 7}, 138 (1976).

\bibitem{balian} R. Balian and C. Bloch, Ann. Phys. {\bf 67}, 229
(1972); M. Brack and R.K. Bhaduri, {\it Semiclassical Physics},
Addison--Wesley, Reading, MA (1997).

\bibitem{strut} V.M. Strutinsky, Sov. J. Nucl. Phys. {\bf 3}, 449
(1966); Nucl.  Phys. {\bf A 95}, 420 (1967); {\it ibid }{\bf A 122}, 1
(1968); M. Brack {\it et al.}, Rev. Mod. Phys. {\bf 44}, 320 (1972).

\bibitem{bohigas} O. Bohigas {\it et al.}, Phys. Rep.  {\bf 223}, 43
(1993); O. Bohigas {\it et al.}, Nucl. Phys. A {\bf 560}, 197 (1993);
S. Tomsovic and D. Ullmo, Phys. Rev. E {\bf 50}, 145 (1994);
S.D. Frischat and E. Doron, Phys. Rev. E {\bf 57}, 1421 (1998).

\bibitem{gian} O. Bohigas {\it et al.}, Phys. Rev. Lett. {\bf 52}, 1
(1984).

\bibitem{mehta} M.L. Mehta, {\it Random Matrices}, Academic Press
Inc., Boston, 1991.

\bibitem{hans}  T. Guhr, A. M\"{u}ller--Groeling and
H.A. Weidenm\"{u}ller, Phys. Rept. {\bf 299}  189 (1998).

\bibitem{bulwirz} A. Bulgac, A. Wirzba, nucl--th/0102018.

\bibitem{lor} C.P. Lorenz {\it et al.},
Phys. Rev. Lett. {\bf 70}, 379 (1993).

\bibitem{chris} G. Baym, H.A. Bethe and C.J. Pethick, Nucl. Phys. {\bf
A175}, 225 (1971); C.J. Pethick and
D.G. Ravenhall, Ann. Rev. Nucl. Part. Sci. {\bf 45}, 429 (1995).

\bibitem{heiselberg} H. Heiselberg {\it et al.}, Phys. Rev. Lett.
{\bf 70}, 1355 (1992).

\bibitem{bul2} A. Bulgac, P. Magierski, unpublished

\bibitem{rosch} A. Rosch, {\em Quantum--coherent transport of a heavy
particle in a fermionic bath} (Shaker--Verlag, Aachen, 1997);
A. Rosch, T. Kopp, Phys. Rev. Lett. {\bf 75}, 1988 (1995);
Phys. Rev. Lett. {\bf 80}, 4705 (1998);

\end{thebibliography}
\end{document}